\documentclass[]{revtex4}
\usepackage{graphics}
\begin{document}
\title{Continuous variable quantum key distribution using polarization encoding
and post selection}
\author{Stefan Lorenz$^1$, Natalia Korolkova$^{2,1}$, and Gerd Leuchs$^1$}
\address{(1) Institute of Optics, Information and Photonics, University Erlangen-N\"urnberg, 
Staudtstra{\ss}e 7/B2, D-91058 Erlangen, Germany, (2) School of Physics and Astronomy, University of St. Andrews, North Haugh, St. Andrews KY16 9 SS, UK}
\date{\today}
%

%
\begin{abstract}
We present an experimental demonstration of a quantum key distribution protocol using coherent polarization states. Post selection is used to ensure a low error rate and security against beam splitting attacks even in the presence of high losses. Signal encoding and readout in polarization bases avoids the difficult task of sending a local oscillator with the quantum channel. This makes our setup robust and easy to implement. A shared key was established for losses up to 64\%.
\end{abstract}
\keywords{03.67.-a, 03.67.Dd, 42.50.Lc, 42.50.-p}
\maketitle
\section{Introduction}
\label{Introduction}
The first quantum cryptography protocols used measurements of noncommuting observables of single photons to establish a secret shared key between two parties \cite{GIS02}. By evaluating the errors in the shared key, it is possible to spot a potential eavesdropper and estimate his knowledge about the key. As there are no reliable, deterministic and fast single photon sources available at the moment, most implementations of such single photon systems use weak coherent pulses instead. Due to the problem of multi-photon components of coherent states in those systems \cite{BRA00c}, the effective amplitude has to be very low, which again impairs the performance. A second drawback is the lack of fast and efficient single photon detectors, whereas bright light photoreceivers work with nearly unit quantum efficiency at high speeds. Thus, a new number of quantum cryptography systems employing coherent states have been proposed \cite{BEN92,HUT95,NAM03,BAR03}. To establish a shared key when the state is attenuated---single photons are either completely lost or transmitted---special  techniques have been proposed \cite{SIL02,GRO02b}. Reverse reconciliation used in \cite{GRO03} demonstrated for the first time the robustness of continuous variable systems against losses of more than 3dB, but it requires strict one way communication and relies on interferometric stability for the transmission of a local oscillator beam. For low transmission losses a cryptography system using post selection \cite{SIL02} of quadrature measurements was presented in \cite{HIR00,HIR03}, using also a separate phase reference pulse. In this work, we demonstrate experimentally for the first time that post selection of coherent states can also establish a shared key in the presence of high losses. In contrast to other experiments \cite{HIR00,GRO03}, which encode information in the phase of coherent states, we use polarization encoding. This dispenses with the need of a separate local oscillator beam, and increases detection efficiency, making our system more suitable for practical applications.\\
\section{Post selection}
\label{Post_selection}
A coherent state with amplitude $\alpha$ is an eigenstate of the annihilation operator $\hat{a}$ and can be represented as an expansion in the Fock basis with photon number $n=\langle \hat{n} \rangle = \langle \hat{a}^\dagger \hat{a} \rangle $ \cite{GLA63b}: 
\begin{equation}
|\alpha\rangle=e^{-\frac{{|\alpha|}^2}{2}}\sum\frac{\alpha^n}{{(n!)}^{\frac{1}{2}}}|n\rangle.
\end{equation}
As formulated by the Heisenberg uncertainty principle, a coherent state occupies a certain area in phase space, such that two coherent states $|{+}\alpha\rangle=e^{i0}|\alpha\rangle$ and $|{-}\alpha\rangle=e^{i\pi}|\alpha\rangle$ exhibit an overlap 
\begin{equation}
f=e^{-2|\alpha|^2}.
\label{Gleichung_Overlap}
\end{equation}
As a result, a measurement of the quadrature amplitude $x= \langle \hat{a}^\dagger + \hat{a} \rangle$ cannot discrimiate deterministically between $|{+}\alpha\rangle$ and $|{-}\alpha\rangle$ for small $|\alpha|$, as measuring of $|{+}\alpha\rangle$ may yield results that could have been produced by a measurement of a $|{-}\alpha\rangle$ state and vice versa. When a sender (Alice) prepares randomly one of the two non-orthogonal states, and a receiver (Bob) guesses which state it was by measuring $x$, his error probability $p_e=Prob(x<0 \mid  |{+}\alpha\rangle )+Prob(x>0 \mid |{-}\alpha\rangle )$ is directly linked to the amplitude $|\alpha|$ and the resulting overlap $f$. The shared information between Alice and Bob, $I_{AB}$, can then be determined. It depends on Alice's amplitude $|\alpha|$, Bob's amplitude measurement result $x$ and the transmission $\eta$ of the channel between Alice and Bob \cite{SIL02,HOR03,NAM03}:
\begin{equation}
I_{AB}=1+p_e \log_2 p_e + (1-p_e) \log_2 (1-p_e).
\end{equation}
We now consider a beam splitting attack, where a potential eavesdropper (Eve) splits off part of the signal which is lost in the channel with transmitivity $\eta$, and transmits the rest with a hypothetical lossless channel, such that her presence is undetectable. Eve can then make measurements on her part of the signal, e.g. an amplitude measurement like Bob. A crucial point in the experiment is that for coherent states Eve's and Bob's individual measurement outcomes are independent within their respective probability distributions. Thus Eve may obtain inconlusive results while Bob is quite confident about the state Alice prepared, and vice versa. Alice and Bob can determine the error probability and mutual information $I_{AB}$ for each single event after the measurement. The average information Eve can get depends on the amplitude $\alpha$ Alice prepares and Eve's portion of the signal (in this case $\sqrt{1-\eta}$), giving a mutual information of $I_{AE}$ between Alice and Eve. For a given channel transmission $\eta$ and state overlap governed by $|\alpha|$ a threshold $t$ for Bob's measurement can be given so that all his results with $|x| > t$ meet the condition $I_{AB}>I_{AE}$. The shared knowledge between Alice and Bob about such events is larger than that shared between Alice and Eve, allowing a secret key to be distilled. The process of sorting out those events which are favourable for Alice and Bob after the data have been recorded is called post selection \cite{SIL02}.\\
\section{Polarization states}
\label{Polarization_states}
The measurement of conjugate quadratures (e.g. amplitude and phase) of an optical mode normally requires a separate phase reference (local oscillator). Our system utilizes the quantum polarisation of a two mode coherent state, providing its own built in strong reference field. The quantum Stokes operators are used throughout this work to describe the quantum polarisation. They are defined in analogy to the classical Stokes parameters \cite{KOR02}. If one defines the annihilation operators $\hat{a}_{x/y}$ for two orthogonal polarization modes the Stokes operators read:
\begin{eqnarray}
\hat{S}_{0} &= \hat{a}^{\dagger}_{x} \hat{a}_{x} + \hat{a}^{\dagger}_{y}
\hat{a}_{y}, \quad
\hat{S}_{1} &= \hat{a}^{\dagger}_{x} \hat{a}_{x} - \hat{a}^{\dagger}_{y}
\hat{a}_{y}, \nonumber \\
\hat{S}_{2} &= \hat{a}^{\dagger}_{x} \hat{a}_{y} + \hat{a}^{\dagger}_{y}
\hat{a}_{x}, \quad
\hat{S}_{3} &= i(\hat{a}^{\dagger}_{y} \hat{a}_{x} - \hat{a}^{\dagger}_{x}
\hat{a}_{y}).
\end{eqnarray}
Their commutator and the corresponding uncertainty relations (with $V$ denoting the variance)
\begin{equation}
\left[ \hat{S}_k,\hat{S}_l\right]  = 2i\varepsilon_{klm}\hat{S}_m, \quad k,l,m=1,2,3
\label{Gleichung_Unschaerfe}
\end{equation}
\begin{equation}
V_{k} V_{l} \geq|\varepsilon_{klm}\langle\hat{S}_m\rangle|^{2}
\end{equation}
ensure that no two operators can be measured simultaneously with certainty, as long as the third is nonzero. This behaviour is similar to quadrature operators, with the exception that the size of the uncertainty area of two observables depends on the mean value of the third. 
In the experiment, the $S_1$ component is chosen to be large, which corresponds to an almost completely horizontally polarized light beam with $S_2$ and $S_3$ as noncommuting observables. The state overlap in $S_2$ and $S_3$ can be utilized analogous to the state overlap in field quadratures.
The Stokes operators are measured by direct detection \cite{KOR02}. The mode with high photon number $\hat{a}_x$ is used as a phase reference to determine the photon number in the dark mode $\hat{a}_y$ of orthogonal polarization. The $S_2$ and $S_3$ components can be measured by applying appropriate phase shifts between $\hat{a}_x$ and $\hat{a}_y$ and using a balanced photodetector. Note that in conventional homodyne detection, the signal and the local oscillator are in two spatially separated modes. Thus the spatial overlap and the phase stability limit the efficiency of such a setup, while our polarization setup has perfect spatial overlap and a stable relative phase by default without active control.\\
\section{Protocol}
\label{Protocol}
\begin{figure}
  \includegraphics{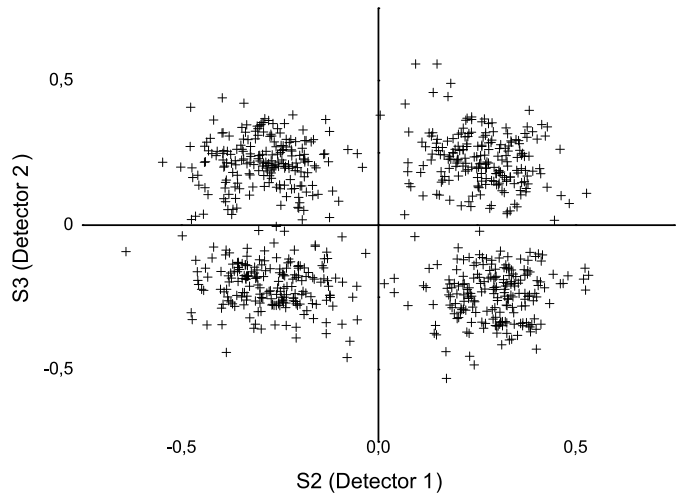}
\caption{Plot of possible Q-function measurement results for Bob. Alice produces four coherent states with either positive or negative $S_2$ and $S_3$ polarization. In the experiment, the state overlap is high, thus the states cannot be distinguished. In this figure, the overlap is low for better visualization.} 
\label{Messung_S2S3_Verteilung}
\end{figure}

The key distribution protocol works with a BB84-type prepare and measure strategy. By small modulations of the $S_1$ polarized cw beam, four coherent states with slightly positive (negative) $\langle\hat{S_2}\rangle$ and $\langle\hat{S_3}\rangle$ are produced. Fig. \ref{Messung_S2S3_Verteilung} shows possible measurement outcomes for Bob, when he uses a 50:50 beam splitter to measure $S_2$ on one half, and $S_3$ on the other half of the beam (see  also Fig. \ref{Schema_Bob_DDet}).
Alice prepares randomly one of these four states and Bob chooses randomly a measurement basis out of $S_2$ and $S_3$. By assigning a bit value "1" ("0") to a positive (negative) measurement result, a shared key can be established. Then Bob discards all results that did not exceed the post selection threshold. He then announces his measurement bases through the public channel so that Alice knows Bob's measurement result with high probability. In the case of vanishing overlap between the states, this procedure is deterministic and hence insecure. An eavesdropper may discriminate between all four states and launch an intercept/resend type attack without being noticed. By using states with a considerable overlap, the error probability for Eve increases, while Bob and Alice can post select favourable events.\\
\section{Experiment}
\label{Experiment}
Our experimental apparatus consists of independent setups for Alice and
Bob which are separated by roughly 30~cm. Alice controls the laser source and the state preparation. Bob performs the polarization measurements on the states he receives from Alice.
\subsection{Alice}
\begin{figure}
\resizebox{0.40\columnwidth}{!}{%
\includegraphics{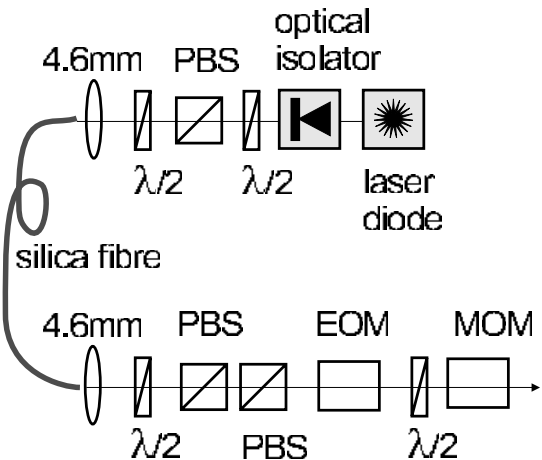}}
\caption{Schematical view of Alice's setup.} 
\label{Schema_Alice}
\end{figure}

Her schematical setup is shown in Fig. \ref{Schema_Alice}.
As a light source a commercial diode laser is used (TOPTICA DL100).
The diode provides up to 40~mW cw optical power and is wavelength stabilised to 810~nm. After mode cleaning of the laser light by a standard telecom fibre (SMF 1528) it is polarized in $S_1$ direction using two polarising beam splitters for improved purity. The $S_3$ modulation is
achieved by an electro-optical modulator (EOM). The $S_2$ modulation of the beam is done by a magneto-optical modulator (MOM) which uses the Faraday effect of a magneto-optically active glass rod (Moltech MOS-04). The overall attenuation of the sender setup was adjusted to give a cw output power of 0.5~mW of $S_1$ polarized light. By controlling the applied modulation voltage on the EOM and the current through the MOM coil, the polarization amplitude in the $S_3$ and $S_2$ direction can be adjusted continuously. Each single event in the experiment is represented by a 500~$\mu$s long interval of modulation on the continuous $S_1$ beam.
\begin{figure}
\resizebox{0.4\columnwidth}{!}{%
\includegraphics{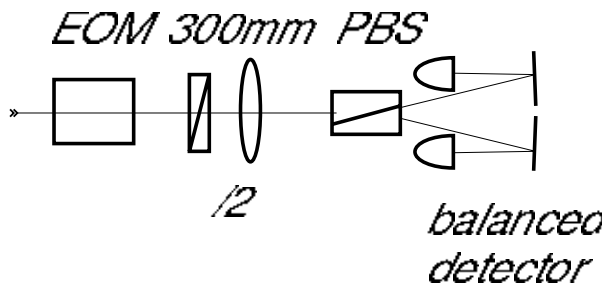}}
\caption{Schematical view of Bob's setup.} 
\label{Schema_Bob}
\end{figure}
\subsection{Bob}
At the receiving station Bob measures either the $S_2$ or $S_3$ displacement. The incoming beam polarization is adjusted according to \cite{KOR02} by a half wave plate and a second EOM. To switch between $S_2$ and $S_3$ the EOM changes from "no retardation" to "quarter wave retardation". The beam is focussed by a lens through a high quality calcite Wollaston polarizer with an extinction coefficient higher than $10^6$. The two resulting
beams are each reflected by a low loss mirror onto silicon PIN photodiodes (Hamamatsu
S3883). The detectors show a linear behaviour up to 1~mW incident power. With an incident power of 250~$\mu$W, the electronic noise is more than 10~dB below the signal noise for modulation frequencies above 20~kHz. The signal is recorded by a fast digitizing oscilloscope (Tektronix TDS~420) and transferred to a computer for Bob's data processing.\\
\section{Results}
\label{Results}
\begin{figure}
\resizebox{0.4\columnwidth}{!}{%
\includegraphics{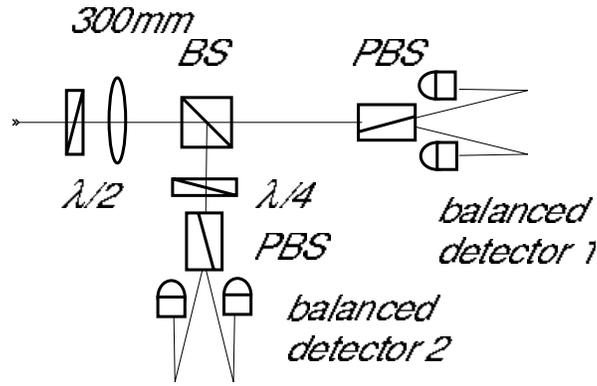}}
\caption{Schematical view of Bob's setup, using two detection setups to measure simultaneously $S_2$ and $S_3$ with a loss of 50\%. If the quarter wave plate is removed, $S_2$ can be measured on both halves of the original beams.} 
\label{Schema_Bob_DDet}
\end{figure}

\begin{figure}
\resizebox{0.6\columnwidth}{!}{%
\includegraphics{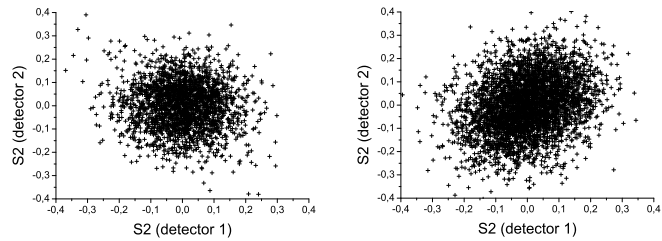}}
\caption{Signal divided by a 50:50 beam splitter and measured by two $S_2$ detectors. Left: no $S_2$ modulation of the signal. Right: small $S_2$ modulation, correlations of the two detected signals result in elliptical shape of the scatter plot. As detector 1 and detector 2 have slightly different gains, a small horizontal ellipticity is introduced in both graphs.} 
\label{Plot_DDet_Korrelationen}
\end{figure}

To show the independence between single results of the two measurement distributions in the case of a beam splitting attack, the signal is divided by a 50:50 splitter and measured by two Stokes detector setups (see Fig. \ref{Schema_Bob_DDet}).
When both detectors are set to measure in the same basis (in this case $S_2$), an unmodulated $S_1$ polarized coherent state gives rise to independent Gaussian distributions in each detector. The results of detector 1 and detector 2 are uncorrelated, as can be seen in Fig. \ref{Plot_DDet_Korrelationen} (left). For a $S_2$ modulation with low amplitude, the plot (Fig. \ref{Plot_DDet_Korrelationen}, right) shows a slight ellipticity, revealing small correlations between the signals of the two detectors. 
Although the two parties share common information it is not possible to deduce the sign of the measurement result of one detector from the outcome of the other detector with certainty. A potential eavesdropper who uses detector 2 cannot infer the results of detector 1, even though she has measured 50\% of the signal. Note that the setup of Fig. \ref{Schema_Bob_DDet} was also used with one $S_2$ and one $S_3$ detector to produce Fig. \ref{Messung_S2S3_Verteilung}.\\ 
In the QKD experiments, only a single detector with electro-optical basis switching (see Fig. \ref{Schema_Bob}) was used. The losses due to non-unity photodiode efficiency, limited EOM transmitivity and optical imperfections in the detector were treated as if they were transmission channel losses. This conservative point of view implies that Eve could manipulate Bob's receiver and increase its efficiency while sending more imperfect states at the same time, gaining additional information. With an additional attenuator between Alice and Bob, the overall transmission levels were set to either 79\% or 36\%. The modulation was adjusted to give an average coherent amplitude of $| \alpha | =0.6$ in the dark mode $\hat{a_y}$, yielding an overlap of $f=0.5$ according to Eq.(\ref{Gleichung_Overlap}). 
\begin{figure}
\resizebox{0.5\columnwidth}{!}{
\includegraphics{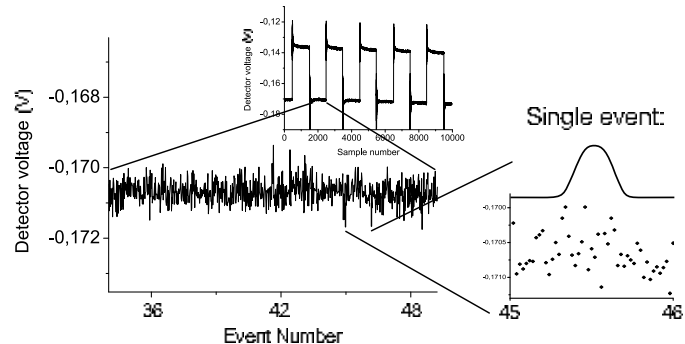}}
\caption{A typical oscilloscope trace of the detector signal with constant basis. Owing to the large overlap the signal modulation is not apparent in the graph. The inset shows the whole trace, with visible jumps due to basis switching by the EOM. One single event, with modulation pattern (solid line) and oscilloscope samples (dots) is shown on the right.} 
\label{Bild_Samples_EOMH}
\end{figure}
A typical readout of the oscilloscope is shown in Fig. \ref{Bild_Samples_EOMH}. The main graph shows eleven intervals of constant measurement basis. The upper inset shows the oscilloscope trace for a period of 0.1~$s$ corresponding to 200 measurement intervals. The abrupt changes of the signal are due to the basis switching by Bob's EOM. Alice's modulation pattern for one single event, as well as Bob's oscilloscope samples are shown on the right. By integrating over the oscilloscope samples in the center of the event where Alice applied a modulation, Bob retrieved a measurement result, while the samples at the beginning and the end of the event interval served as calibration data.\\
Bob then assigned a bit value to each measurement. After that, all events where Bob's measurement value was below the post selection threshold were discarded. Table \ref{Tabelle_Raten_T} shows the relevant parameters for the case of low loss as well as for high loss. All rates are given in bits per second. While the raw bit rates are nearly equal for both scenarios, a higher post selection threshold was used in presence of 64\% loss, to ensure a considerable information advantage $I_{AB}-I_{AE}$ for the selected bits. Thus the bit rate after post selection is much lower for the high loss case. Note also that the errors in Bob's key drop dramatically after post selection, so that a subsequent error correction with standard procedures is now possible. The higher error probability in the high loss case originates from the increased overlap of the states (cf. Eq. (\ref{Gleichung_Overlap})) as the coherent amplitude $\alpha$ decreased with transmission losses.\\
The bit rate after error correction with the established \textsc{Cascade} protocol \cite{BRA94} is given for the optimum block length and 5 passes of the protocol. The potential final bit rate can be calculated from the information advantage and the error corrected bit rate. Bits needed to authenticate the public channel are not taken into account. In both loss scenarios it was possible to produce a nonzero bit rate of shared bits.\\
Note that, in contrast to the reverse reconciliation method used in \cite{GRO03}, the error correction does not take place on the raw key (with approx. 1kbit/sec) but on the post selected key (with approx. 400bit/sec and 150bit/sec respectively), thus needing less capacity on the classical channel. Also, we achieved a key exchange with direct reconciliation in the presence of 64\% loss, corresponding to a $-4.4dB$ quantum channel transmittivity.\\
\begin{table}
\begin{tabular}{ccc}
\hline\noalign{\smallskip}~                            ~&~21\% loss~&~64\% loss~\\\noalign{\smallskip}\hline\noalign{\smallskip}
~raw bit rate in bit/s           ~&~ 1069    ~&~ 1096    ~\\
~errors before post selection~&~ 22.0\%  ~&~ 27.3\%  ~\\
~bit rate after post selection   ~&~ 415     ~&~ 165     ~\\
~errors after post selection ~&~ 6.0\%   ~&~ 7.6\%   ~\\
~post selection threshold t ~&~ 1.0     ~&~ 2.3     ~\\
~information advantage       ~&~ 0.76    ~&~ 0.49    ~\\
max. bit rate after error correction&~ 249     ~&~ 80      ~\\
max. bit rate after priv. amp.&~ 189 ~&~ 39 ~\\
\noalign{\smallskip}\hline
\end{tabular}
\caption{Experimental key generation parameters for high and low losses. The raw rate, rate after post selection and information advantage as well as the error fraction are experimental results. The maximum rates after error correction and privacy amplification are theoretical predictions for the availabe data. The post selection threshold is given in units of the coherent amplitude $\alpha$.}
\label{Tabelle_Raten_T}
\end{table}
\section{Conclusions}
\label{Conclusions}
The system can be further improved e.g. in the receiver design. If one uses a setup similar to that used to generate the plot in Fig. \ref{Messung_S2S3_Verteilung} (see Fig. \ref{Schema_Bob_DDet}), Bob's detector would not require active basis switching, like in some single photon systems, e.g. \cite{KUR02b}. The theoretical upper bound for detection speed is then the bandwidth of the balanced detector, which can be very high compared to photon counters in single photon experiments. When using polarization dispersion compensation, the scheme can also be adapted to fibre transmission lines by replacing the 810~nm coherent laser source with a 1.5~$\mu$m laser.\\
To conclude, we have demonstrated a quantum key distribution system that relies on readily available coherent
states, continuous variable measurements and post selection to generate a shared key between Alice and Bob. The states are
polarization encoded, ensuring fast modulation,
and perfect mode overlap in Bob's homodyne detector. There is no need to send a separate local oscillator along with the quantum channel. The system is robust against losses of more than 50\% and does not need special reconciliation techniques which require strict one way error correction protocols such as in \cite{GRO03}. Its implementation is straightforward and can be used e.g. in free space communication with high efficiency and key exchange rate.
\section{Acknowledgments}
This work was supported by the Federal Ministry of Education and Research (BMBF/VDI) under FKZ:13N8016. The authors would like to thank U.~Andersen, J.~Schneider and A.~Berger for technical assistance and for help with the preparation of this manuscript.

\bibliographystyle{amsplain}
\bibliography{kryptobib}

\end{document}